\documentclass[12pt,twocolumn]{article}

\usepackage{epsfig}
\usepackage[numbers]{natbib}
\usepackage{graphicx}
\usepackage{color}
\usepackage[final]{pdfpages}
\usepackage{fullpage}
\begin{document}
\bibliographystyle{abbrvnat}

\title{Non-standard morphological relic patterns in the cosmic microwave background}
\author{Joe Zuntz\footnote{Astrophysics Group, Oxford University \& Astrophysics Group, University College London. email: jaz@astro.ox.ac.uk}, James P. Zibin\footnote{
Department of Physics \& Astronomy, University of British Columbia}, Caroline Zunckel\footnote{Astrophysics and Cosmology Research Unit, University of Kwazulu-Natal}, Jonathan Zwart\footnote{Department of Astronomy, Columbia University}}
\date{1 April 2011}
\maketitle

\begin{abstract}
Statistically anomalous signals in the microwave background have been extensively studied in general in multipole space, and in real space mainly for circular and other simple patterns.  In this paper we search for a range of non-trivial patterns in the temperature data from WMAP 7-year observations.  We find a very significant detection of a number of such features and discuss their consequences for the essential character of the cosmos.
\end{abstract}

\section{Introduction}
\label{Introduction} The most significant information contained in maps of the cosmic microwave background (CMB) and large-scale structure has for the most part been contained solely in their two-point statistics: the power spectra and angular correlation functions of the respective fields.  But it has long been recognized that important information can be lost if we exclusively follow this approach.

One area of fertile or at least extensive research has been searches for non-gaussian statistics of the CMB.  The bispectrum and trispectrum of Wilkinson Microwave Anisotropy Probe (WMAP) and other data have been analyzed for consistency with gaussianity, with varying results \citep{ng1,ng2,ng3,ng4,ng5} and even single point statistics beyond the variance can provide strongly convincing evidence of non-gaussian behavior \citep{smoot}.

There have also been searches for real-space (morphological) patterns in the CMB.  Various models suggest that there should be `circles-in-the-sky' present on very large scales: large annular regions of correlated or enhanced background temperature.  These patterns might be generated by non-trivial cosmic topologies \citep{top1,top2,top3,top4,top5}, or by more exotic pre-big bang models of cosmological history.  There have been claims of a highly significant detection of concentric circles in the latter context in WMAP and other data \citep{penrose}, but they have been contradicted by other analyses \citep{MossScottZibin,WehusEriksen,Hajian}, which have suggested that such signals could be caused by systematic error.

There is little work in the literature concerning non-circular relic patterns in the CMB, apart from the pioneering search for triangular patterns in \citep{MossScottZibin}.  Elliptical patterns have been considered, but it has been noted that ellipses are really just stretchy circles \citep{CitationNeeded}.

In this article we search for other morphologies in primary CMB data by correlating patterns with the WMAP 7-year W-band data \citep{wmap-maps}.  Section \ref{Introduction} contains the standard description of things you already know.  We copied and pasted some text from a previous paper into section \ref{Methodology}.  In section \ref{patterns} we discuss the theoretical underpinning of the relic patterns for which we search.  In section \ref{results} we describe results of searches for such patterns.  

Throughout this article we use the convention that happiness and morality are positive and sadness and evil negative.  We assume the standard mind-the-gap tube announcement.
\section{Methodology}
\label{Methodology}

\subsection{Posterior statistics \& morphology}
\begin{figure}[htbp]
\begin{center}
\includegraphics[scale=0.4]{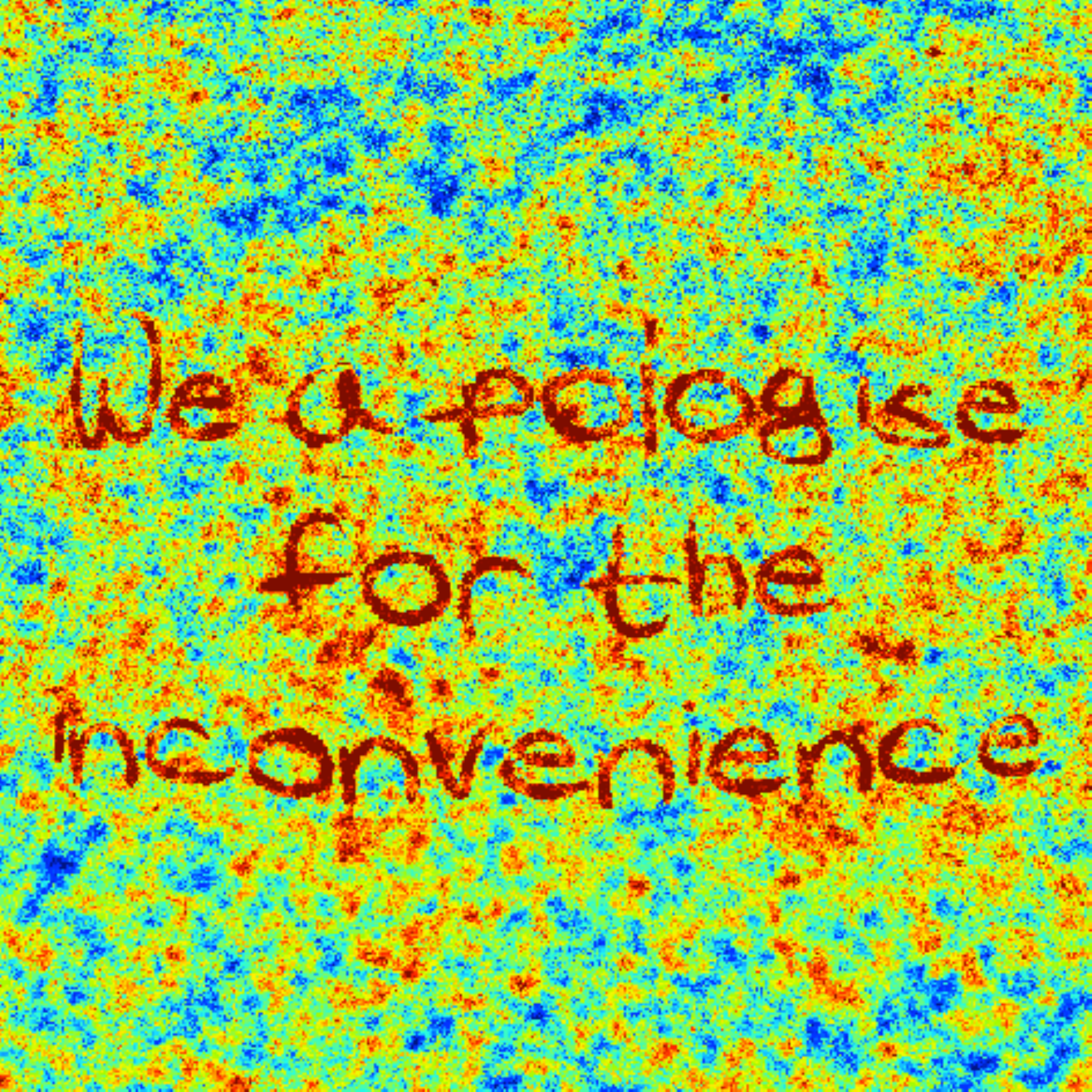}
\caption{An example of a CMB map for which the application of posterior statistics would not be wholly unreasonable.}
\label{apologize}
\end{center}
\end{figure}

It has been stated that selection of one's CMB statistic \emph{a posteriori} is on some level bad science \citep{wmap-anomalies}.  We must object to this on two counts.  On a practical level cosmic variance limits us; if we were to follow this principle exactly we would be permitted no more hypotheses about the CMB after the first WMAP data release.   Second, there is clearly some level of significance of a statistic so-selected that overcomes any objection.  For example, if the primary cosmic microwave background anisotropies had the words \emph{We apologise for the inconvenience} written in 300$\mu$K hot letters at the Galactic north pole (see Figure \ref{apologize}) then we would not be persuaded that this has no significance, even if no one suggested it in advance (though in fact they did: \citep{ZooLetters}).

\subsection{Methods}
It has often been noted by theorists that the conceptual simplicity and beauty of a theory is as important as such quotidian concerns as accurate fits to data.  We adopt this philosophy now and apply it to the data analysis method we use in this paper.  Our analysis will be the computational equivalent of a theoretical toy model: simple, but we do not want to imply we are not clever enough to do a better one, just too busy.

We generate template images of the patterns discussed in section \ref{patterns}, and convert them to {\sc Healpix} \citep{healpix} maps at resolution $N_\mathrm{side}=512$, initially centered at the north pole.  We compare these template maps to co-added WMAP 7-year maps of the W-band microwave sky as follows: For each pixel on a lower resolution $N_\mathrm{side}=64$ map, we rotate the positions of those pixels `hit' in the template image so that their north points to that pixel.  We then take the covariance of these pixels to the WMAP pixels at the newly rotated positions as the new low-resolution pixel value.  We exclude pixels in the WMAP temperature analysis mask.  We normalize each pixel by the mean temperature of pixels in a disc of fixed radius around it, so that simple CMB hotspots are not detected.

\section{Search patterns \& theoretical basis}
\label{patterns} There is some previous literature on whether the information content of the CMB could contain messages of universal, observer-independent significance \citep{ScottZibin,HsuZee}.  We can characterize this approach as an extended application of the Copernican principle, which states that our perspective on the Universe should be in some sense typical.  As an alternative, we can adopt the approach taken in the study of so-called \emph{void} cosmologies by otherwise mostly sensible people \citep{ZibinMossScott,CliftonFerreiraZuntz,ZZZZ}.  In that approach we drop this assumption and posit that the Milky Way and Earth are in some central cosmological location.  

We can take a similar approach here; the claim that any information content in the CMB must be universal might be termed the Cultural Copernican Principle.  As an alternative we can look for morphological patterns that do have parochial cultural relevance.  Analyses of the string-theory landscape typically show that anthropically controlled parameters should take the most extreme values they can while still resulting in the formation of observers \citep{landscape}.   

Applying a similar argument\footnote{Or at least one that is approximately as meaningful.}, we may suppose that any parochial information in the CMB is nonetheless as universal as possible.  Since the only cultural sample we have is our own, we must therefore look for relatively popular local symbols that frequently appear in ways that seem to suggest the communication of information.  
\subsection{Patterns}
Table 1 shows the morphological patterns for which we search in this paper.  There is usually an element of subjectivity in the choice of such patterns; we have made strenuous efforts to avoid this by consciously being very very fair.

The most universal information encoding devised so far is the unicode character set \citep{unicode}, which attempts to include all human communication symbols.  The majority of the symbols we use are thus drawn from this set.  We have also attempted to include symbols with meanings that are as far as possible opposites, so that we may draw more general statistical conclusions about the nature of the CMB. We must consider both increments and decrements in temperature; but note that negative pattern A is not the same as positive pattern B, for example.

For each pattern, we should as faithful bayesians provide a model prior which describes the degree of credence we assign to the presence of the pattern, \emph{before we look at the data}, or if this is not possible, temporarily forgetting the data.  Fortunately this is rather simple in this case: each pattern can either be present or not present, so the probability is 0.5.

\includepdf{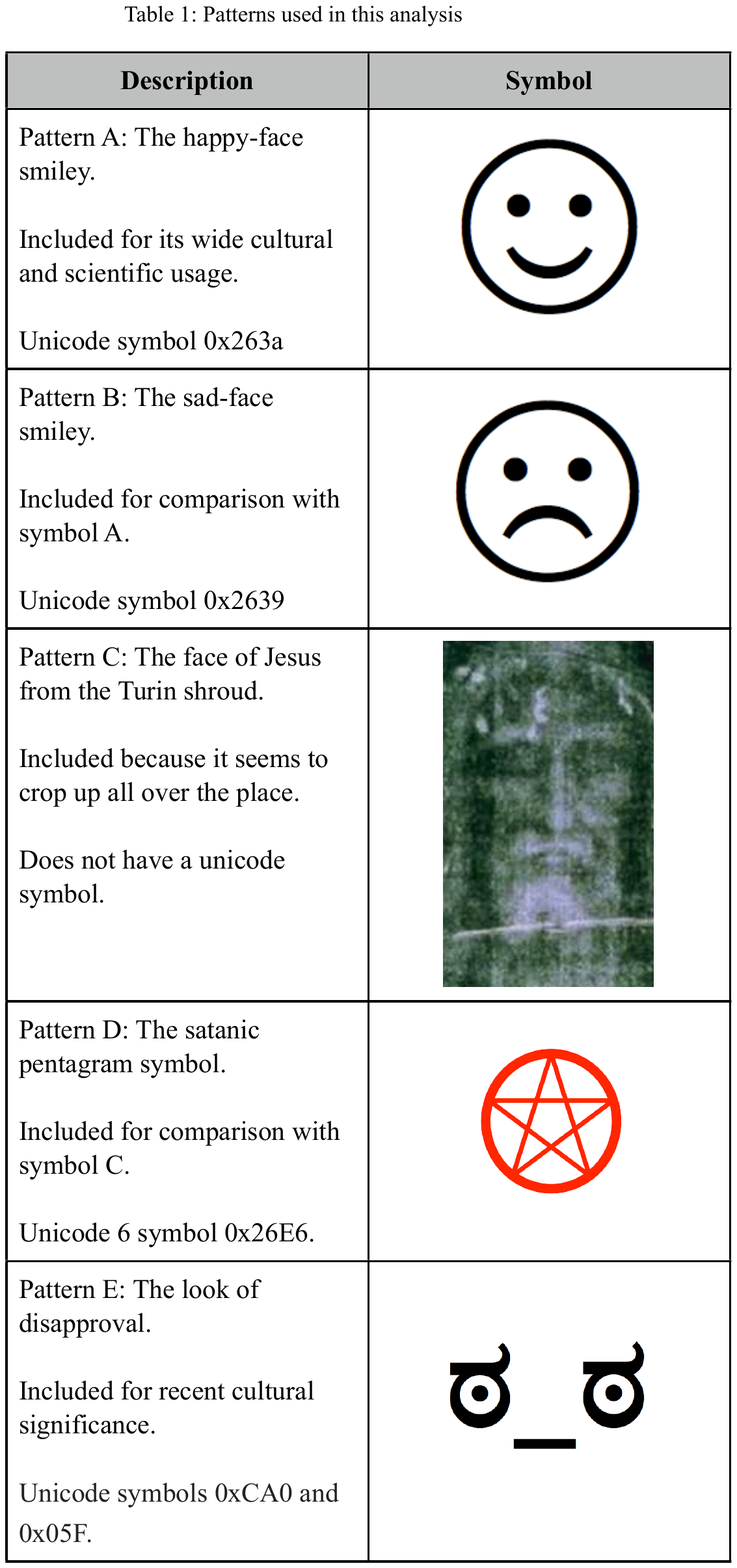}

\section{Results}
\label{results}
\subsection{Relative Significance}

\begin{figure}[htbp]
\begin{center}
\includegraphics[scale=0.4]{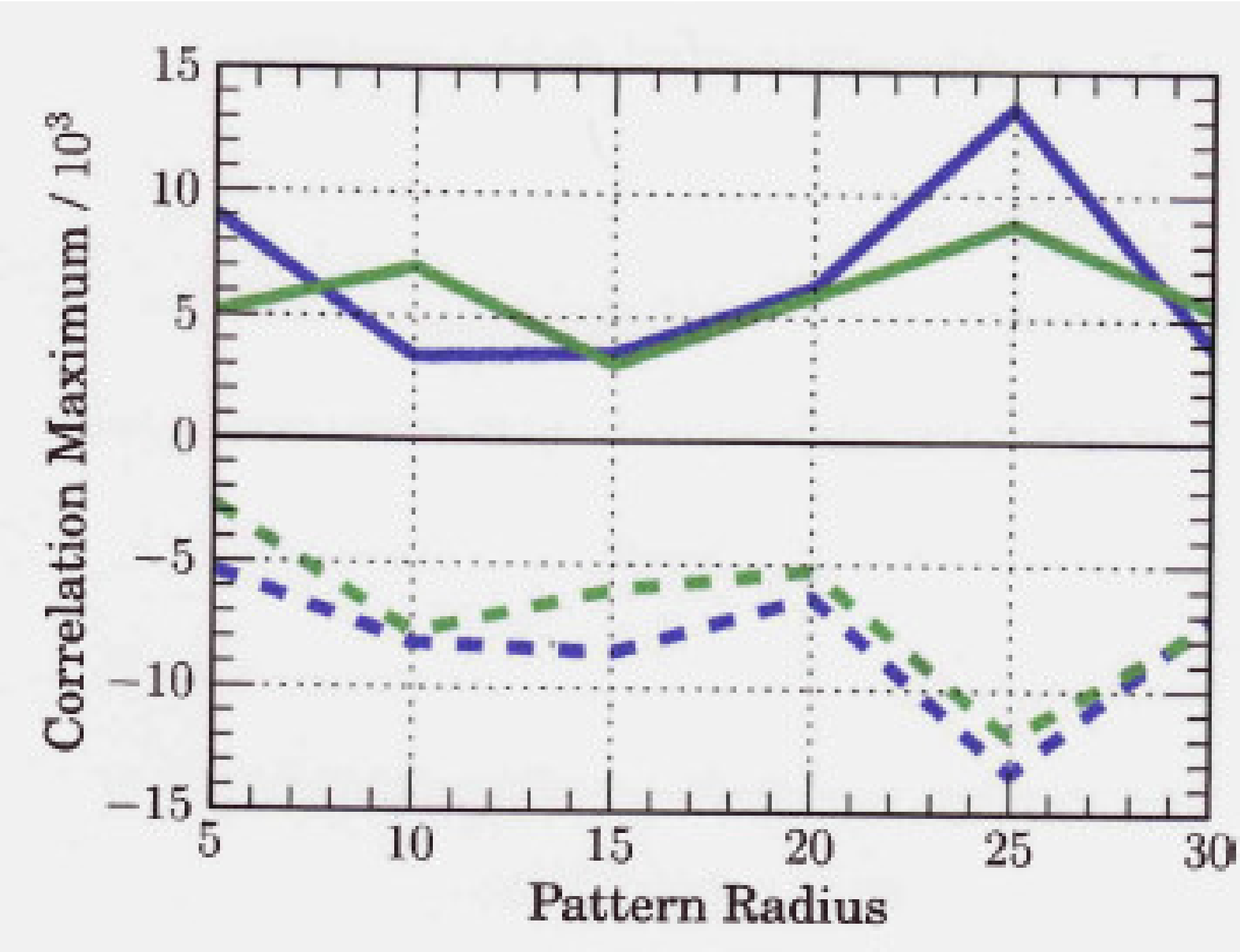}
\caption{Detection levels of patterns A (happy face; green) and B (sad face; blue) with radius.  The maximum signal at each radius is the solid line and the minimum is the dashed line.}
\label{happy-sad-figure}
\end{center}
\end{figure}

\begin{figure}[htbp]
\begin{center}
\includegraphics[scale=0.4]{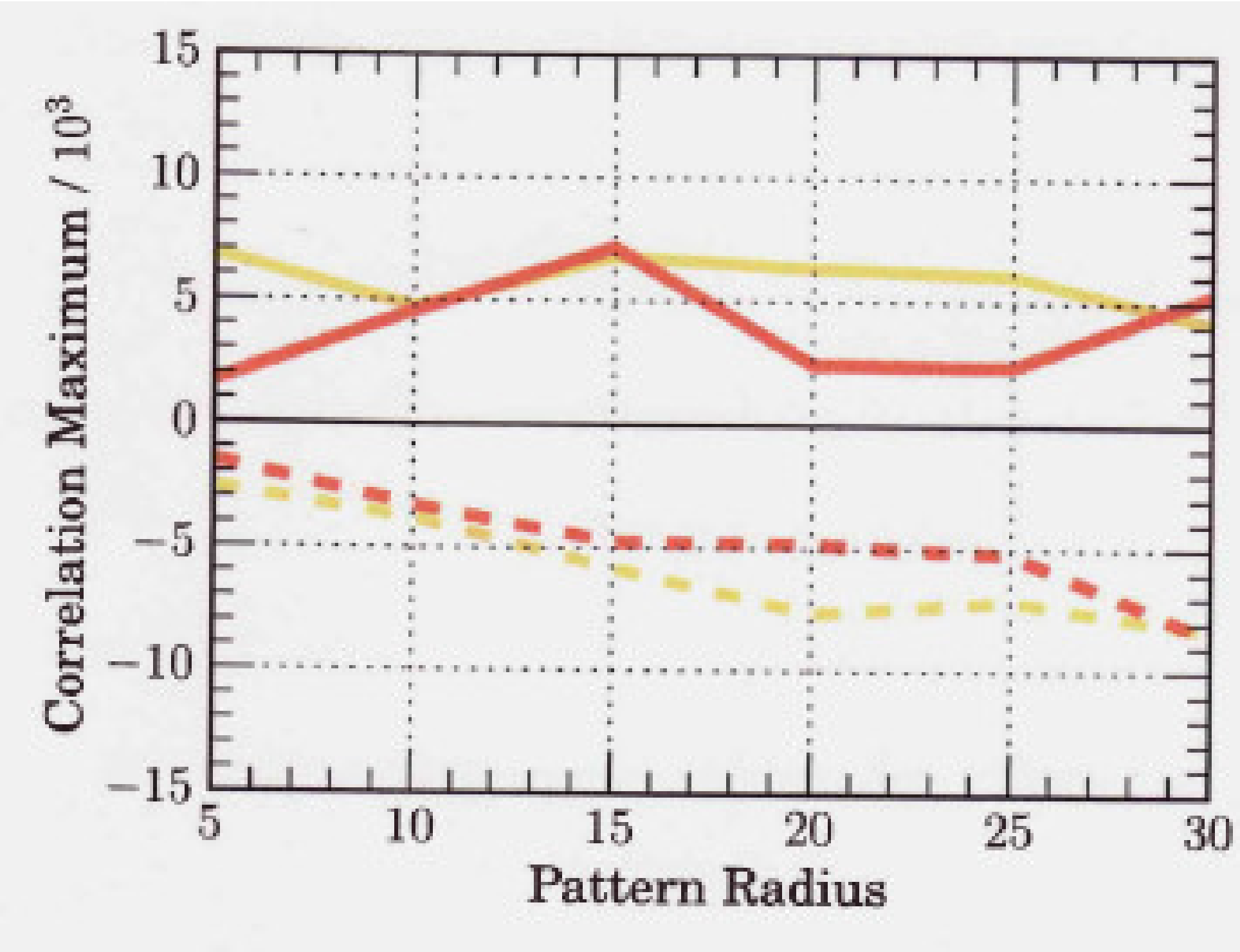}
\caption{Detection levels of patterns C (face of Jesus; yellow) and D (pentagram; red) with radius.}
\label{good-evil-figure}
\end{center}
\end{figure}

\begin{figure}[htbp]
\begin{center}
\includegraphics[scale=0.4]{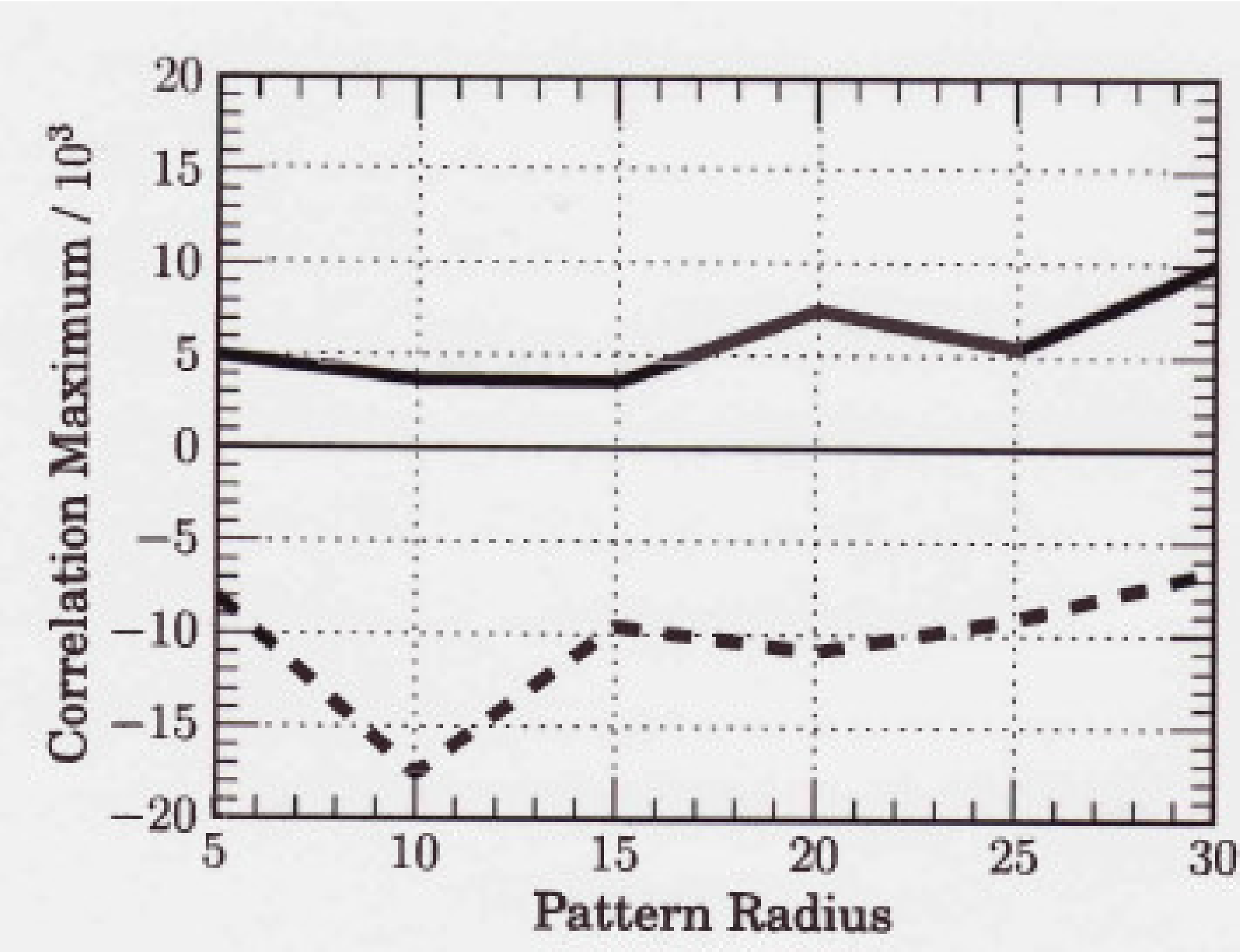}
\caption{Detection levels of patterns E (look-of-disapproval) with radius. }
\label{look-figure}
\end{center}
\end{figure}

The level at which we detect patterns as a function of their radii are shown in Figures \ref{happy-sad-figure}, \ref{good-evil-figure} and \ref{look-figure} for patterns (A,B), (C,D) and E respectively.  We can immediately pick out several key features.

First, the pattern B curve in Figure \ref{happy-sad-figure} has more power than pattern A; this is particularly prominent for the negative regime.  This indicates that the CMB is somewhat sad.

Similarly, pattern C is more prominent than pattern D in Figure \ref{good-evil-figure}, indicating that the CMB signal is a holy one; this is further evidence against the axis-of-evil phenomenon reported at large scales \citep{axisevil}.

\begin{figure}[htbp]
\begin{center}
\includegraphics[scale=0.6]{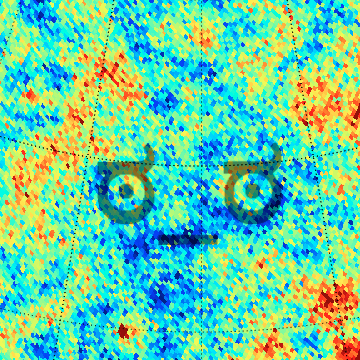}
\caption{A detection of pattern E, the look of disapproval.}
\label{look-inplace}
\end{center}
\end{figure}

The strongest peak in all the patterns is in the negative regime in Figure \ref{look-figure}, for pattern E, at a radius of $10\deg$, corresponding to an angular multipole $\ell \approx 18$.  This indicates that the CMB is strongly disapproving at those scales.  We might speculate about the cause of this disapproval; it seems to be associated with the famed `cold spot'.  The location of the strong disapproval cold spot is shown in Figure \ref{look-inplace}.

\subsection{Absolute significance}
We can, as above, easily compare two different signals in the CMB to generate a relative significance of two detected patterns.  But it is perhaps more instructive to compute an absolute significance level so that we can state a familiar `sigma-number' that everyone thinks they understand.

We therefore run our pattern search code on simulated data for comparison.  There are of course various levels of sophistication we might use for these simulations.  The most complete simulation would include foregrounds, WMAP scan strategies, and correlated noise.  We might use the method recommended in \citep{penrose2}, where cosmological CMB signal in the maps is excluded and only noise is considered.  It is clear, however, that this runs the risk of over-estimating the probability of finding a pattern by chance; the more conservative choice is clearly to analyze a map with neither signal nor noise using the same pipeline.

We find that the likelihood of finding any such patterns in these simulations is zero; the absolute detection significance of our discoveries is therefore approximately $\infty \sigma$. 

\section{Conclusions}
\label{conclusion} We have shown that common methods of CMB circle-searching can be straightforwardly extended to non-circular patterns, and have applied such a method.  In doing so, we have been able to characterize previously unmeasured statistics about the microwave background: its mood, characterized by the difference in detection significance between patterns A and B, and its moral fiber, characterized by the difference between patterns C and D.  We find that the CMB is sad, good and disapproving, which is perhaps a bittersweet conclusion.  

There are of course several other similar statistical CMB measurements we could make: its political persuasion, for example, or its gender or sexuality\footnote{The authors have always conceived of the CMB as an elderly lesbian Tea-Partier; this perhaps says more about them than it.}; we leave these to future work.  We expect our findings to inspire theoretical studies to elucidate the underpinnings of these unanticipated aspects of the CMB.  Indeed, we suggest that such efforts begin immediately, since it is known that the CMB anisotropies are time-dependent \citep{ZibinMossScott2}, and so follow-up observational studies of the CMB may indicate that the effects we describe are only transient.  If subsequent reanalyses fail to find the same results as those found here then that is perhaps the reason.

We have highlighted in this paper the deficiency of two-point statistics alone in studying the CMB or other fields.  It seems clear that this habit, which has arisen because we have only two eyes and so can see only two data points at once, needs to be augmented with new methods.

\emph {Acknowledgments} Many people contributed ideas and help for this manuscript; for some reason they wished to remain anonymous.  The exception was Olaf Davis, who we thank for useful conversations.


\end{document}